\documentclass[aps,pra,superscriptaddress,twocolumn]{revtex4-2}
\usepackage{amssymb,bm}
\usepackage{graphicx}
\usepackage{amsmath}
\usepackage{epstopdf}
\usepackage{float}
\allowdisplaybreaks
\usepackage[T2A]{fontenc}
\usepackage[utf8]{inputenc}
\usepackage[english]{babel}

\begin{document}
 \begin{titlepage}

\title{Non-stationary Aharonov-Bohm effect}
\author{A.~I.~Milstein}\email{a.i.milstein@inp.nsk.su}
\affiliation{Budker Institute of Nuclear Physics of SB RAS, 630090 Novosibirsk, Russia}
\affiliation{Novosibirsk State University, 630090 Novosibirsk, Russia}
\author{I.~S.~Terekhov}\email{i.s.terekhov@gmail.com}
\affiliation{School of Physics and Engineering, ITMO University, 197101 St. Petersburg, Russia}

\date{\today}

\begin{abstract}
The non-stationary Aharonov-Bohm effect (scattering of electron in the field of a narrow solenoid with alternating current) is considered.  Using the eikonal approximation,  the wave function of  electron,  the differential and total scattering cross sections are found. Unlike the case of direct current, the total cross section in the case of alternating current turns out to be finite. An oscillating asymmetry in the differential scattering cross section is discovered. The possibility of experimental observation of the effect is discussed.	
\end{abstract}

\maketitle
 \end{titlepage}
\section{Introduction}
The scattering of an electron on a narrow long solenoid with a direct current, the Aharonov-Bohm effect \cite{AB1959}, is one of the striking manifestations of quantum mechanics. The change of the scattering cross section with a change of the magnetic flux in the region inaccessible to the electron motion is surprising. Many papers have been devoted to the study of the Aharonov-Bohm effect (see, for example, the review~\cite{PT1989}). A distinctive feature of the Aharonov-Bohm effect is the topology of  space accessible to the electron motion (punctured line in a space). If a direct current is passed through the solenoid, then, by virtue of Maxwell   equations, the electric and magnetic fields outside the solenoid will be equal to zero.

However, if the current in the solenoid is alternating, the electromagnetic fields will be non-zero in the accessible to the electron motion region.  In this case, the topology of the space accessible to the electron motion is the same as that in the stationary case,  so that  the influence of  alternating magnetic field flux inside the solenoid on the electron motion outside the solenoid (the non-stationary Aharonov-Bohm effect) will be somewhat similar to the stationary effect, but will also have some differences.  The non-stationary Aharonov-Bohm effect has been discussed in many papers with contradictory results (see, for example, \cite{LYG1992, ADC2000, GNS2011, SV2013, JJ2017, CM2019}). However, none of these papers calculated such experimentally important quantities as the differential and total scattering cross sections. Our work is devoted to the study of these quantities.

\section{Eikonal approximation and the cross section of non-stationary Aharonov-Bohm effect}

Let us consider an infinitely long solenoid with radius $a$ and the number $n$ of turns per unit length, through which a current $I(t)=I_0\,\cos(\omega t)$ is passed. We denote
\begin{align}
	\Phi_0=\dfrac{4\pi^2a^2}{c}\,nI_0\,,
\end{align}
where $c$ is the speed of light. At $\omega=0$ this quantity coincides with the magnetic field flux through the solenoid. We are interested in the limit $a\rightarrow 0$ at a fixed value of $\Phi_0$. Solving Maxwell equation, we find the vector-potential  $\bm A(\bm r,t)$, the magnetic field $\bm B(\bm r,t)$ and the electric field $\bm E(\bm r,t)$ outside the solenoid,
\begin{align}
	&\bm A(\bm r,t)=  [\bm\nu\times\bm n] \,A(r,t)\,,\,\, \bm B(\bm r,t)=  \bm\nu\, B(r,t)\,, \nonumber\\
	&\bm E(\bm r,t)=  [\bm\nu\times\bm n] \,E(r,t)\,,\nonumber\\
	 &A(r,t)=\Phi_0\dfrac{k}{4}[J_1(kr)\sin(\omega t)-N_1(kr)\cos(\omega t)]\,,\nonumber\\
	&B(r,t)=\Phi_0\dfrac{k^2}{4}[-J_0(kr)\sin(\omega t)+N_0(kr)\cos(\omega t)]\,,\nonumber\\
	&E(r,t)=-\Phi_0\dfrac{k^2}{4}[J_1(kr)\cos(\omega t)+N_1(kr)\sin(\omega t)]\,,
\end{align}
where $\bm\nu$ is a unit vector directed along the $z$-axis and parallel to the solenoid axis, $\bm n=\bm r/r$, $\bm r=(x,y,0)$, $r=\sqrt{x^2+y^2}$, $k=\omega/c$, $J_l(x)$ is the Bessel function, $N_l(x)$ is the Neumann function.

For $kr\ll 1$ we have
\begin{align}
	&A(r,t)=\dfrac{\Phi_0}{2\pi r}\cos(\omega t)\,,\nonumber\\
	&B(r,t)=k^2\dfrac{\Phi_0}{2\pi}\ln(kr)\cos(\omega t)\,,\nonumber\\
	&E(r,t)=k\dfrac{\Phi_0}{2\pi r}\sin(\omega t)\,.
\end{align}	
 For $kr\gg 1$ the asymptotics of fields are
\begin{align}
	&A(r,t)=\Phi_0\sqrt{\dfrac{k}{8\pi r}}\cos(\omega t-kr+\pi/4)\,,\nonumber\\
	&B(r,t)=-\Phi_0 k\sqrt{\dfrac{k}{8\pi r}}\sin(\omega t-kr+\pi/4)\,,\nonumber\\
	&E(r,t)=k\Phi_0\sqrt{\dfrac{k}{8\pi r}}\sin(\omega t-kr+\pi/4)\,.
\end{align}
These asymptotics correspond to the radiation field of a solenoid.

We consider a non-relativistic problem, so that we assume that the electron velocity $v$ satisfies the condition $v/c\ll 1$. The Pauli equation for the wave function $\psi(\bm r,t)$ of an electron in the field of a solenoid reads
\begin{align}
	 i\hbar\dfrac{\partial\psi(\bm r,t)}{\partial t}=&\Bigg[ \dfrac{1}{2M}\left(\bm p-\dfrac{e}{c}[\bm\nu\times\bm n]A(r,t) \right)^2\nonumber\\
	 &-\lambda\mu\, B(r,t)\Bigg]\,\psi(\bm r,t)\,,
\end{align}
where $M$, $e$ and $\mu$ are the mass, charge and magnetic moment of  electron, $\lambda=\pm 1$ corresponds to the projection $\pm 1/2$ of  spin  onto the $z$ axis. Since $A(r,t)$ and $B(r,t)$ are periodic functions in $t$ with the period $T=2\pi/\omega$, the wave function satisfies the periodicity condition \cite{Ritus1967,Zeldovich1973},

\begin{align}
\psi(\bm r, t+T)=e^{i\alpha T}\psi(\bm r, t)\,.\label{translEq}
\end{align}

Let us rewrite the Pauli equation in the form
\begin{align}
	 i\hbar\dfrac{\partial\psi(\bm r,t)}{\partial t}=&\Bigg[ \dfrac{\bm p^2}{2M}-\dfrac{e\hbar\,A(r,t)}{Mcr}\,l_z\nonumber\\
	 &+\dfrac{e^2\,A^2(r,t)}{2Mc^2}-\lambda\mu\, B(r,t)\Bigg]\,\psi(\bm r,t)\,,
\end{align}
where $l_z=(xp_y-yp_x)/\hbar$ is the projection operator of  orbital momentum onto the $z$ axis.
Let the momentum of  incident electron be $P$. From the experimental point of view, the momenta $P\gg \hbar k$ and small scattering angles $\theta\ll 1$ are of interest, while the relationship between $q=P\theta/\hbar$ and $k$ can be arbitrary. Under these conditions, the eikonal approximation \cite{LLQM} is valid and we can write  the wave function in the form
\begin{align}
\psi(\bm r,t)=\exp\Bigg\{-i\dfrac{P^2t}{2M\hbar}+i\dfrac{Px}{\hbar}\Bigg\}\,F(\bm r,t)\,.\label{wave_function}
\end{align}
Then,
\begin{align}\label{feq}
	& i\hbar\dfrac{\partial F(\bm r,t)}{\partial t}= -i\dfrac{\hbar P}{M}\,\dfrac{\partial F(\bm r,t)}{\partial x} + \dfrac{eP\,y\,A(r,t)}{Mcr}\,F(\bm r,t)\,,
\end{align}
where we have dropped the terms $\partial^2F/\partial x^2$, $\partial^2F/\partial y^2$ and kept only the terms that contain a large momentum $P$ in the numerator.
This equation is independent of spin orientation.
Using the method of characteristics, we find the solution to Eq.~\eqref{feq},
\begin{align} 
	&F(x,y,t)=\exp\left\{\!-i\dfrac{ey}{ c\hbar}\int\limits_{-\infty}^x\!\! dz \frac{A(x,y,t+(z-x)/v)}{\sqrt{y^2+z^2}} \right\}\,,\label{FuncitonF}
\end{align}	
where $v=P/M$. 
The shift in the last argument of the function $A(x,y,t)$ corresponds to the retardation effect. Note that  the  wave function \eqref{wave_function}, with $F(x,y,t)$ given by \eqref{FuncitonF}, satisfies the condition \eqref{translEq}.

To obtain the scattering cross section, it is necessary to find the limit 
$$S(y,t')=\lim_{x\rightarrow\infty}F(x,y,t)$$
at a fixed value $t'=t-x/v$. Replacing in \eqref{FuncitonF} the upper limit of the integral  with $\infty$, we get
\begin{align} 
	&S(y,t')=e^{-i\Xi(y,t')}\,,\nonumber\\
	& \Xi(y,t')=\dfrac{ey}{ c\hbar}\,\int_{-\infty}^{\infty}\dfrac{dz}{\sqrt{y^2+z^2}}\,A(x,y,t'+z/v)\,,
\end{align}	
Since the field is time-dependent, the particle flux (cross-section $d\sigma$) is also  time-dependent. Following the standard derivation,  see \cite{LLQM},  we obtain
\begin{align} \label{secdif}
	&d\sigma(\theta,t)=\dfrac{P}{2\pi\hbar}\,\left| \int_{-\infty}^{+\infty}dy\, e^{-iqy}\left[1-\,e^{-i\,\Xi(y,t')}\right]\,\right|^2d\theta\,,
\end{align}	
Making the substitutions of variables $|q|y\rightarrow y$, $|q|z\rightarrow z$ and $\omega t'\rightarrow \tau $, we get
\begin{align} \label{EqSigma}
	&d{\sigma}(\theta,\tau)=\dfrac{\hbar}{2\pi P}\,\left| \int_{-\infty}^{+\infty}dy\, e^{-is y}\left[1-\,e^{-i\,\Xi(y,\tau)}\right]\,\right|^2\dfrac{d\theta}{\theta^2}\,,\nonumber\\
	&\Xi(y,\tau)=\dfrac{\zeta  yQ}{ 2 }\,\,\int_{-\infty}^\infty\dfrac{dz}{r}\,[J_1(\zeta r)\sin(\phi)-N_1(\zeta r)\cos(\phi)]\,,\nonumber\\
	&\zeta=\dfrac{k}{|q|}=\dfrac{\hbar k}{P\theta}\,,\quad Q=\dfrac{e\Phi_0}{2 c\hbar}\,,\quad \phi=\tau+\dfrac{c}{v}\,\zeta z\,,\nonumber\\ 
	&r=\sqrt{y^2+z^2}\,,\quad s=\mbox{sgn}\,q\,.
\end{align}	
Recall that $\tau=\omega(t-x/v)$ and the cross section in the two-dimensional case has the dimension of length. Let us show that in the stationary case the expression \eqref{EqSigma} goes over to the known result. To do this, we set $\omega\rightarrow 0$ before the change of variables, and obtain the differential scattering cross section for the stationary Aharonov-Bohm effect \cite{AB1959,LLQM}
\begin{align} 
 d\sigma =\dfrac{2\hbar}{\pi P}\,\sin^2Q\,\dfrac{d\theta}{\theta^2}\,.
\end{align}
For $v/c\ll 1$ and $\zeta\lesssim 1$ the main contribution to the cross section is determined by the integration region $y\ll 1$ and $z\ll 1$. In this case we find the asymptotics $\Xi_0(y,\tau)$ of the function $\Xi(y,\tau)$,
\begin{align} \label{Xi0}
	&\Xi_0(y,\tau)=Q\,\mbox{sgn}(y)\,\cos\tau\,\exp(-|y|/\vartheta)\,,\nonumber\\
	& \vartheta= \dfrac{v}{c\zeta}=\dfrac{\theta}{\theta_0}\,,\quad \theta_0=\dfrac{\hbar\omega}{2E}\,.
\end{align}
Using integration by parts and the asymptotics \eqref{Xi0}, we write the expression for the cross section in the form
\begin{align} \label{difG12}
	&d{\sigma}(\theta,\tau)=\sigma_0\,\Big| s\, G_1(\vartheta,\tau)+\dfrac{1}{\vartheta}\,G_2(\vartheta,\tau) \,\Big|^2{d\vartheta}\,,\nonumber\\
	&G_1(\vartheta,\tau)=\int_{0}^\infty dy\,\sin[\Xi_1(y,\tau)]\,\sin(\vartheta y) \,,\nonumber\\
	&G_2(\vartheta,\tau)=\int_{0}^\infty dy\,\Xi_1(y,\tau)\,\sin[\Xi_1(y,\tau)]\,\sin(\vartheta y) \,,\nonumber\\
	&\sigma_0=\dfrac{2\hbar}{\pi P\theta_0}=\dfrac{2v}{\pi\omega}\,,\quad \Xi_1(y,\tau)=Q\,e^{-y}\,\cos\tau\,.
\end{align}	
The function $G_1(\vartheta,\tau)$ changes sign when replacing $\cos\tau\rightarrow- \cos\tau$, while $G_2(\vartheta,\tau)$ does not change sign. For convenience, in Eq.~\eqref{difG12} we have passed from the angle $\theta$ to $\vartheta$.

If $\vartheta\gg 1$ ($1\gg\theta\gg \theta_0$), then
\begin{align} \label{as1}
	&  d{\sigma}(\vartheta,\tau) =\sigma_0\,\sin^2\left(Q\cos\tau\right)\,\dfrac{d\vartheta}{\vartheta^2}\,.
\end{align}	
For $\vartheta\ll 1$ ($\theta\ll\theta_0$), we have
\begin{align}\label{as2} 
	&  d{\sigma}(\vartheta,\tau) =\sigma_0\,\left|\int_{0}^{Q|\cos\tau|}\,(1-\cos y)\,\dfrac{dy}{y}\right|^2d\vartheta\,.
\end{align}	
The asymptotics \eqref{as1} and \eqref{as2} for the cross sections are independent of  $s=\mbox{sgn}(q)$, since either the function $G_1(\vartheta,\tau)$ contributes to the cross section for $\vartheta\gg 1$, or $G_2(\vartheta,\tau)$ for $\vartheta\ll 1$. However, for $\vartheta\sim 1$ an asymmetry arises due to the interference of the functions $G_1(\vartheta,\tau)$ and $G_2(\vartheta,\tau)$. This asymmetry will be an alternating function of $\tau$ and can be observed by measuring the non-stationary Hall effect (oscillating voltage in the transverse direction).

\begin{figure}[h!]
	\centering
	\includegraphics[width=0.8\linewidth]{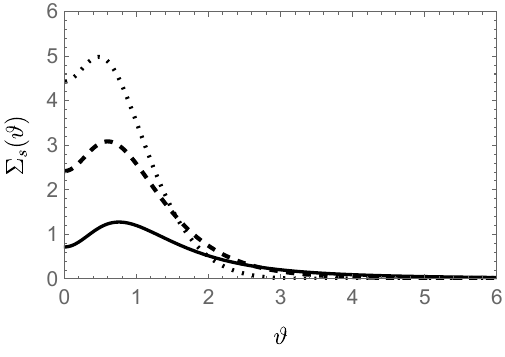}\\
		\includegraphics[width=0.8\linewidth]{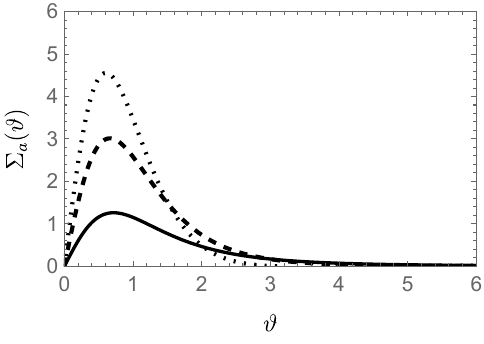}
	\caption{Dependence of $\Sigma_{s}(\vartheta,\tau)= \sigma_0^{-1}d\sigma_{s}(\vartheta,\tau)/d\vartheta$ (top plot) and $\Sigma_{a}(\vartheta,\tau)= \sigma_0^{-1}d\sigma_{a}(\vartheta,\tau)/d\vartheta$ (bottom plot) on $\vartheta$ for  $f=2$ (solid curve), $f=3$ (dashed curve) and $f=4$ (dotted curve), $f=Q\cos\tau$.}
	\label{Sigsa}
\end{figure}

Let us represent the cross section $d\sigma(\vartheta,\tau)$ in \eqref{difG12} in the form
\begin{align} \label{difsa}
	&d\sigma(\vartheta,\tau)=d\sigma_{s}(\vartheta,\tau)+s\,d\sigma_{a}(\vartheta,\tau)\,,\nonumber\\
	&d{\sigma}_{s}(\theta,\tau)=\sigma_0\,\Big[G_1^2(\vartheta,\tau)+\dfrac{1}{\vartheta^2}\,G_2^2(\vartheta,\tau) \,\Big]{d\vartheta}\,,\nonumber\\
	&d{\sigma}_{a}(\theta,\tau)=2\sigma_0\,G_1(\vartheta,\tau)\,G_2(\vartheta,\tau) \,\dfrac{d\vartheta}{\vartheta}\,.
\end{align}
The functions $G_1(\vartheta,\tau)$ and $G_2(\vartheta,\tau)$ depend on the parameter $Q$ and variable $\tau$ only through the combination $f=Q\cos\tau$. Fig.~\ref{Sigsa} shows the dependence of $\Sigma_{s}(\vartheta,\tau)= \sigma_0^{-1}d\sigma_{s}(\vartheta,\tau)/d\vartheta$  and $\Sigma_{a}(\vartheta,\tau)= \sigma_0^{-1}d\sigma_{a}(\vartheta,\tau)/d\vartheta$ on $\vartheta$ for  several values of $f$.

\begin{figure}[h!]
	\centering
	\includegraphics[width=0.8\linewidth]{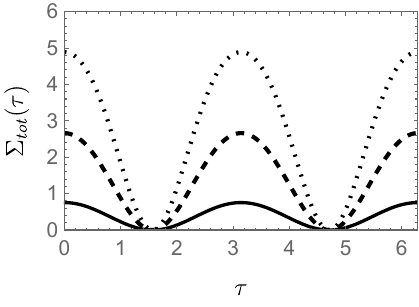}
	\caption{Dependence of $\Sigma_{tot}(\tau)=\sigma(\tau)/\sigma_0$ on $\tau $ for $Q=1$ (solid curve), $Q=2$ (dashed curve) and $Q=3$ (dotted curve).}
	\label{Sigtot}
\end{figure}

It follows from the  asymptotics obtained that for alternating current the total scattering cross section $\sigma(\tau)$  is finite, in contrast to the case of direct current. The main contribution to $\sigma(\tau)$ is given by the region of scattering angles $\theta\sim\theta_0$ ($\vartheta\sim 1$). We have from \eqref{secdif} for $v/c\ll 1$,
\begin{align} \label{sectot1}
	&\sigma(\tau)=\pi\sigma_0\,\int_{0}^{+\infty}dy\,\left[1-\cos\Xi_1(y,\tau)\,\right] \nonumber\\
	&=\pi\sigma_0\,\int_{0}^{Q|\cos\tau|}\,(1-\cos y)\,\dfrac{dy}{y}\,. 
\end{align}	
Naturally, there is no asymmetry in $\sigma(\tau)$. 
The dependence of $\Sigma_{tot}(\tau)=\sigma(\tau)/\sigma_0$ on $\tau $ is shown in 
Fig.\ref{Sigtot} for several values of $Q$.

\begin{figure}[h!]
	\centering
	\includegraphics[width=0.8\linewidth]{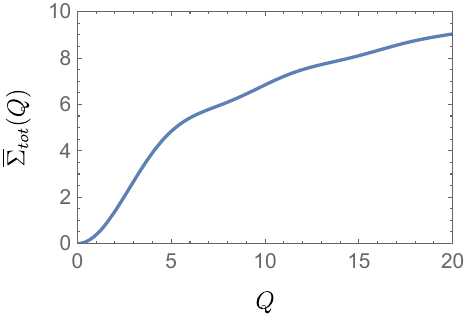}
	\caption{Dependence of  $\overline{\Sigma}_{tot}=\overline{\sigma}(Q)/\sigma_0$ on $Q$.}
	\label{Sigtotav}
\end{figure}

The total cross section increases with $Q$ and  is well approximated at $Q\gg 1$ by the formula
$$\sigma(\tau)\approx \pi\sigma_0\,\ln(1+2Q|\cos\tau|)\,.$$
The total cross section averaged over time may also be of interest,
\begin{align} \label{sectot2}
	&\overline{\sigma}(Q)=\int_0^{2\pi}\sigma(\tau)\,\dfrac{d\tau}{2\pi}=\pi\sigma_0\,\int_{0}^{1}
	\,\left[1-J_0(Qy)\,\right]\dfrac{dy}{y}\,. 
\end{align}	
The dependence of $\overline{\Sigma}_{tot}(Q)=\overline{\sigma}(Q)/\sigma_0$  on $Q$  is shown in Fig.~\ref{Sigtotav}.
The time-averaged cross section grows logarithmically with increasing $Q$ for large $Q$.

\section{Conclusion}
The non-stationary Aharonov-Bohm effect is investigated in the non-relativistic approximation. Using the eikonal approach, the differential and total scattering cross sections are found. In contrast to the stationary Aharonov-Bohm effect, the total cross section in the non-stationary case is finite. The main contribution to the scattering cross section comes from the angles $\theta\sim \theta_0=\hbar\omega/E\ll 1$. The differential scattering cross section contains an asymmetry with respect to the replacement $q\rightarrow -q$. This asymmetry has a maximum at $\theta\sim \theta_0$ and decreases rapidly at $\theta\gg \theta_0$ and $\theta\ll \theta_0$. The asymmetry in the cross section can be observed as the non-stationary Hall effect, and the total scattering cross section manifests itself in oscillations of the total electron flux.

\section*{Acknowledgement}
The work of Ivan Terekhov was financially supported by the ITMO Fellowship Program.

\end{document}